\documentclass[usenatbib,useAMS]{mn2e}

\usepackage{graphicx}
\usepackage{psfrag}

\newcommand{\SA}{subseismic approximation}
\newcommand{\AAP}{anelastic approximation}

\newcommand{\BV}{Brunt-V\"ais\"al\"a\ }\newcommand{\vxi}{\vec{\xi}}
\def\Div{\mathop{\hbox{div}}\nolimits}
\newcommand{\disp}[1]{\displaystyle #1}
\newcommand{\lp}{ \left(}
\newcommand{\rp}{ \right)}
\newcommand{\er}{\vec{e}_r}
\newcommand{\dnr}[1]{\frac{d  #1}{dr}}
\newcommand{\na}{ \vec{\nabla} }
\newcommand{\noi}{ \noindent }
\newcommand{\vg}{\vec{g}}
\def\Ln{\mathop{\hbox{ln}}\nolimits}
\newcommand{\beq}{\begin{equation}}
\newcommand{\eeq}{\end{equation}}
\newcommand{\eeqn}[1]{\label{#1}\end{equation}}
\newcommand{\eps}{\varepsilon}
\newcommand{\YL}{ Y^m_\ell }
\newcommand{\khi}{\chi}
\newcommand{\greq}{\begin{equation}\left\{ \begin{array}{l}}
\newcommand{\egreq}{\end{array}\right. \end{equation}}
\newcommand{\egreqn}[1]{\end{array}\right. \label{#1}\end{equation}}
\newcommand{\beqan}{\begin{eqnarray}}
\newcommand{\eeqan}[1]{\label{#1}\end{eqnarray}}
\newcommand{\eq}[1]{(\ref{#1})}
\newcommand{\vi}{\vec{v}}
\newcommand{\vu}{\vec{u}}
\newcommand{\od}[1]{\mbox{${\cal O}\lp #1\rp$}}
\newcommand{\lc}{ \left[}
\newcommand{\rc}{ \right]}

\title[More concerning the anelastic and subseismic approximations]{More
concerning the anelastic and subseismic approximations for low-frequency
modes in stars}

\author[M. Rieutord and B. Dintrans]{Michel Rieutord$^{1,2}$ and 
Boris Dintrans$^{1,3}$\\
$^1$Laboratoire d'Astrophysique de Toulouse, Observatoire
Midi-Pyr\'en\'ees, 14 avenue E. Belin, 31400 Toulouse, France\\
$^2$Institut Universitaire de France\\
$^3$Nordic Institute for Theoretical Physics, Blegdamsvej 17, DK-2100
Copenhagen, Denmark}

\pagerange{\pageref{firstpage}--\pageref{lastpage}}
\pubyear{2001}

\begin{document}

\bibliographystyle{mn2e}

\maketitle

\label{firstpage}

\begin{abstract}
Two approximations, namely the \SA\ and the \AAP, are presently used to
filter out the acoustic modes when computing low frequency modes of a
star (gravity modes or inertial modes). In a precedent paper (Dintrans
\& Rieutord 2001), we observed that the \AAP\ gave eigenfrequencies much
closer to the exact ones than the \SA. Here, we try to clarify this
behaviour and show that it is due to the different physical approach
taken by each approximation: On the one hand, the \SA\ considers the low
frequency part of the spectrum of (say) gravity modes and turns out to
be valid only in the central region of a star; on the other hand, the
\AAP\ considers the \BV frequency as asymptotically small and makes no
assumption on the order of the modes. Both approximations fail to
describe the modes in the surface layers but eigenmodes issued from the
\AAP\ are closer to those including acoustic effects than their
subseismic equivalent.

We conclude that, as far as stellar eigenvalue problems are concerned,
the \AAP\ is better suited for simplifying the
eigenvalue problem when low-frequency modes of a star are considered,
while the \SA\ is a useful concept when analytic solutions of high order
low-frequency modes are needed in the central region of a star.
\end{abstract}


\section{Introduction}

When considering the low-frequency modes of a star, namely gravity modes or 
inertial modes, the compressibility of the fluid is often a side effect
in the determination of eigenfrequencies and eigenmodes;
in other words, the dynamics of these modes may be simplified by
neglecting the elasticity of the fluid or, equivalently, by filtering
out acoustic modes. This is the aim of the subseismic and anelastic
approximations; the resulting equations for eigenmodes are much simpler
than the original ones and very useful when dealing with the low
frequency oscillations of rotating stars \cite[e.g.][]{DR00}.

Recently, we compared these two approximations (\cite{DR01} referred to
as paper I hereafter). We found that in the two cases which we analysed, namely
two polytropes, the \AAP\ performed much better than the \SA. We attributed
this behaviour to an inconsistency of the \SA\ but our argument turns
out to be not
general and \cite{Smeyers01} showed that, for low-frequency high order
modes, the \SA\ gives the first order equations in regions not close to
the surface of the star. These results prompted us to re-examine this
question in order to clarify the origin of the different behaviour
of the two approximations. For this purpose we will focus, in section 2, on two
asymptotic developments: a first one where we use, as \cite{Smeyers01},
the frequency as a small parameter and a second one where we use the \BV
frequency as the small parameter. These asymptotic developments will
prove to be at the origin of each of these approximations and will
allow us to clarify the physics attached to each of them. In section 3,
using the same examples as in paper I, we will compare the approximate
eigenfunctions to their exact counterparts and show the better behaviour
of the \AAP. Our conclusions are drawn in section 4.

\section{The asymptotic equations}

As was shown in I, both approximations imply Cowling's approximation; we
shall therefore neglect the perturbation of the gravitational potential
and will start from the following equations:

\begin{eqnarray}
& & \rho' + \Div ( \rho \vxi ) = 0, \label{eq1} \\ \nonumber \\
& & \disp \omega^2 \vxi = \na \lp \frac{P'}{\rho} \rp - 
\frac{N^2}{\rho g} \delta P \er, \label{eq2} \\ \nonumber \\
& & \disp \delta P = c^2 \delta \rho, \label{eq3}
\end{eqnarray}
where we assumed a time-dependence of the form $\exp (i \omega t)$ and
considered adiabatic oscillations.  $\vxi$ is the displacement; $P'$
and $\rho'$ respectively denote the Eulerian fluctuations of pressure and
density whereas $\delta P$, $\delta \rho$ are their Lagrangian
counterparts; thus we have

\[
\delta P = P' + \dnr{P} \xi_r, \qquad \delta \rho = \rho' + \dnr{\rho} 
\xi_r,
\]

\noi with a pressure gradient satisfying the hydrostatic equilibrium
$dP/dr=-\rho g$. Also, $\rho$ is the equilibrium density, $\vg=-g\er$ the
gravity and $\gamma=(\partial \Ln P / \partial \Ln \rho)_S$ the first
adiabatic exponent. Finally, $c^2$ and $N^2$ respectively denote the
squares of sound speed velocity and \BV frequency such as

\beq
c^2 = \gamma \frac{P}{\rho}, \qquad N^2=g \lp \frac{1}{\gamma} 
\dnr{\Ln P} - \dnr{\Ln \rho} \rp.
\label{def_bv}
\eeq

\subsection{The subseismic view}

As a first exercise we derive the equations verified by the low
frequency gravity modes.

We therefore assume that the frequency reads $\omega=\eps\omega_1$ and that
$\dnr{}$ scales as $\eps^{-1}$  with $\eps\ll 1$ since we focus on high
radial order modes. Developing the dependent variables generically as

\beq f = f_0+\eps f_1 + \eps^2 f_2 + \cdots, \eeqn{expan1}
and using the classical expansion of the variables on the spherical
harmonics,
\[
\vxi (r,\theta,\phi) = \sum_{\ell=0}^{+\infty}\sum_{m=-\ell}^{+\ell}
\xi^\ell_m(r) \YL(\theta,\phi) \er + \khi_m^\ell(r) \na\YL,
\]
and dropping $(\ell, m)$-indices, we find that $P'_0=P'_1=0$,
$\rho'_0=0$, $\xi_0=0$ and that

\greq
\disp \dnr{} (r\khi_0)  = -\frac{N^2}{\omega_1^2} \xi_1, \\ \\
\disp \dnr{} (r^2\xi_1) =   \ell(\ell+1) r\khi_0.
\egreqn{sub1}

A system which is slightly different from the one obtained when setting
$P'=0$ which yields the subseismic equations:

\beqan
\disp &&\dnr{} (r\khi) = \lp 1-\frac{N^2}{\omega^2}\rp \xi, \label{sub2a}\\
\nonumber \\
\disp &&\dnr{} (r^2\xi) =   \ell(\ell+1) r\khi+\frac{g}{c^2}r^2\xi.
\eeqan{sub2}
However, if we use the expansion \eq{expan1} into
(\ref{sub2a}-\ref{sub2}), we recover \eq{sub1}; therefore,
(\ref{sub2a}-\ref{sub2}) contain terms of higher order than \eq{sub1}.

In fact, our expansion \eq{expan1} breaks down near the origin $r=0$ where the
regularity of the solutions (i.e. that $\xi \propto r^{\ell-1}$) is not
insured. This comes from the fact that terms like $f/r$ are no longer
negligible compared to derivatives $df/dr$.

This difficulty is avoided by \cite{Smeyers01} with the use of the
variable $\tau=\frac{\sqrt{\ell(\ell+1)}}{\eps}\int_0^r\frac{N(r')}{r'}dr'$
instead of the radial variable $r$. $\tau$ serves as a fast variable
while $r$, the slow variable, is assumed small compared to the scale of
variation of the \BV or the background density. Using this transformation,
\cite{Smeyers01} shows that \eq{sub2} is verified by the solution at
leading order while
\eq{sub2a} is approximately verified.

Near the surface layers, \cite{Smeyers01} has shown that within this
development, the Eulerian pressure perturbation is no longer negligible
and that the subseismic approximation does not apply.

Thus the subseismic equations govern the oscillations of high
radial order gravity modes in the central parts of the star. No constraint
is imposed to the \BV frequency and the equations to be solved are:

\[ \Div\vxi = \frac{g}{c^2}\xi_r, \qquad \omega^2\vxi=\na\lp\frac{P'}{\rho}\rp
+ N^2\xi_r\er \]

\begin{figure*}
\centerline{\includegraphics[width=8cm,angle=+90]{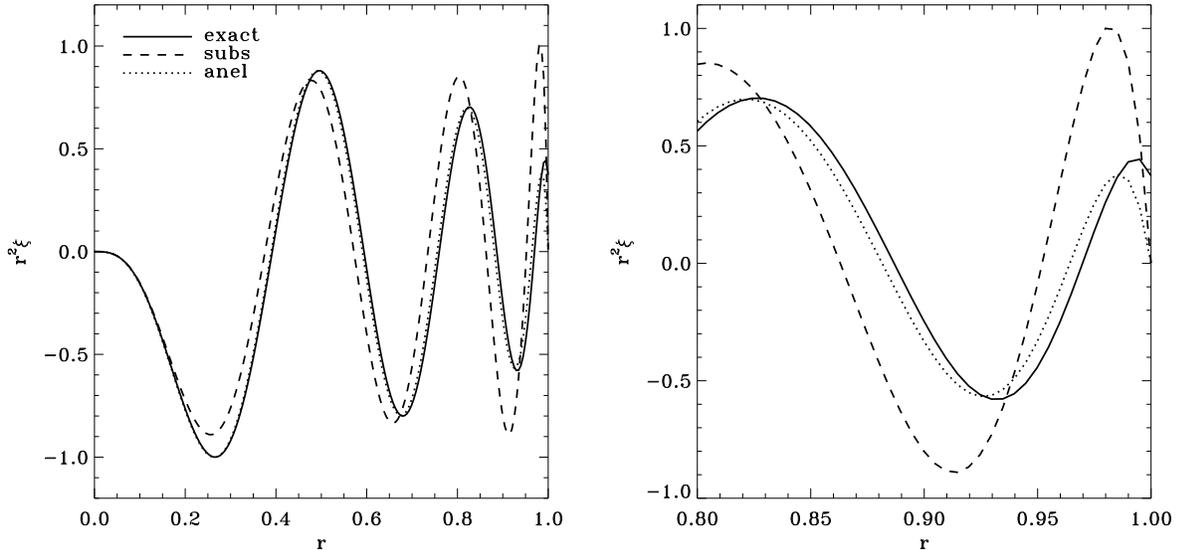}}
\caption[]{Normalized eigenfunctions $\xi$ with $\ell=2$ and $k=5$
for the homogeneous polytrope. The solid line shows the exact solution
($\omega^2\simeq -3.61\times 10^{-2}$) while the dashed and dotted lines
correspond to its subseismic and anelastic approximations, respectively
(with $\omega^2_{\hbox{\tiny subs}}\simeq -3.26\times 10^{-2}$
and $\omega^2_{\hbox{\tiny anel}}\simeq -3.53\times 10^{-2}$). On right,
the surface layers have been magnified.} 
\label{homo} 
\end{figure*}

\subsection{The anelastic view}

Let us now turn to the \AAP. In this case, it is more convenient to
write the equations of motion for the velocity field $\vi$ rather than
the displacement $\vxi$, that is,

\greq
\disp i\omega\rho'+\frac{1}{r^2}\dnr{}(r^2\rho u) - \ell(\ell+1)\frac{\rho
v}{r}=0,\\ \\
\disp i\omega u = -\dnr{}\lp\frac{P'}{\rho}\rp+\frac{N^2}{i\omega\rho g}
(i\omega P' -\rho g u), \\
\disp i\omega rv = - \frac{P'}{\rho}, \\
\disp i\omega (P'-c^2\rho') = -\frac{\rho c^2N^2}{g} u,
\egreq
where we used the following spherical harmonics decomposition for $\vi$
(for clarity, we still dropped in the previous system the $\ell,m$
indices)

\[
\vi (r,\theta,\phi) = \sum_{\ell=0}^{+\infty}\sum_{m=-\ell}^{+\ell}
u^\ell_m(r) \YL(\theta,\phi) \er + v_m^\ell(r) \na\YL.
\]

We now assume that the \BV frequency is vanishingly small; note that as
this quantity often diverges at the surface of the star models, it is
more appropriate to assume that $\omega_N$, the frequency of the lowest
order gravity mode, is vanishingly small. Thus we write

\[ N(r)=\eps n(r),\quad \omega=\eps\omega_1,
\quad P'=P'_0+\eps P'_1+\cdots,\; \ldots
\]
Note that we make no assumption concerning the scale of the perturbations
which may be of order unity. First orders yield the equations

\beqan
&&P'_0=\rho'_0=0, \\
&&\na\cdot\rho\vu_0 = 0, \\
&&i\omega_1u_0=-\dnr{(P'_1/\rho)}-\frac{n^2}{i\omega_1}u_0, \\
&&i\omega_1rv_0=-P'_1/\rho, \\
&&i\omega_1\lp P'_1-c^2\rho'_1\rp = -\frac{\rho n^2c^2}{g}u_0.
\eeqan{aa1}
from which we write the anelastic system

\[ \Div(\rho\vxi)=0,\qquad \omega^2\vxi =\na(P'/\rho)+N^2\xi_r\er \]
As for the subseismic approximation the perturbed equation of state is
eliminated; but from \eq{aa1}, we note that the Eulerian
pressure perturbation is of the same order as $c^2\rho'$.

The subseismic equations can be obtained by just dropping out the
Eulerian fluctuation $P'$ in $\delta P$, in the original equations
(\ref{eq1}-\ref{eq3}); from \eq{aa1}, we see that this is not the case
in the \AAP. On the other hand, the fluctuation of density in the mass
conservation equation can be neglected.

We therefore see that the \AAP\ applies when the \BV frequency is
small compared to the acoustic frequencies but does not impose any
constraint on the scale of the solutions. Near the
surface the anelastic solution differs from the exact solution
because of the different boundary condition: Exact solutions verify
$\delta P=0$, a condition which transforms into $\xi=0$ or $u=0$ for the
approximate solution.

\subsection{Comments}

The foregoing developments show that the \AAP\ applies under rather more
general conditions than the \SA; indeed, the only requirement is the
smallness of the \BV frequency or, in other words, a large separation
between the acoustic spectrum and the gravity spectrum. As this latter
condition is often met in stars we can expect that the \AAP\ performs
better when applied to star models.

Concerning the \SA, it is clear that it can be applied in the central
region of a star but that surface layers should be avoided. Smeyers
introduces the notion of boundary layer to describe the regions where
his asymptotic solutions are valid. However, these boundary layers are
somehow special since their thickness can be comparable
to the radius of the star (in polytropes for instance)\footnote{
Classical boundary layers have a thickness
very small compared to the size of the domain and which tends to zero as
the small parameter is decreased.}.

Broadly speaking, it turns out that the \SA\ has a rather local character
while the \AAP\ has a global one. As eigenvalue problem are global
problems in nature, the \AAP\ should be better suited for these problems.

\begin{figure*}
\centerline{\includegraphics[width=8cm]{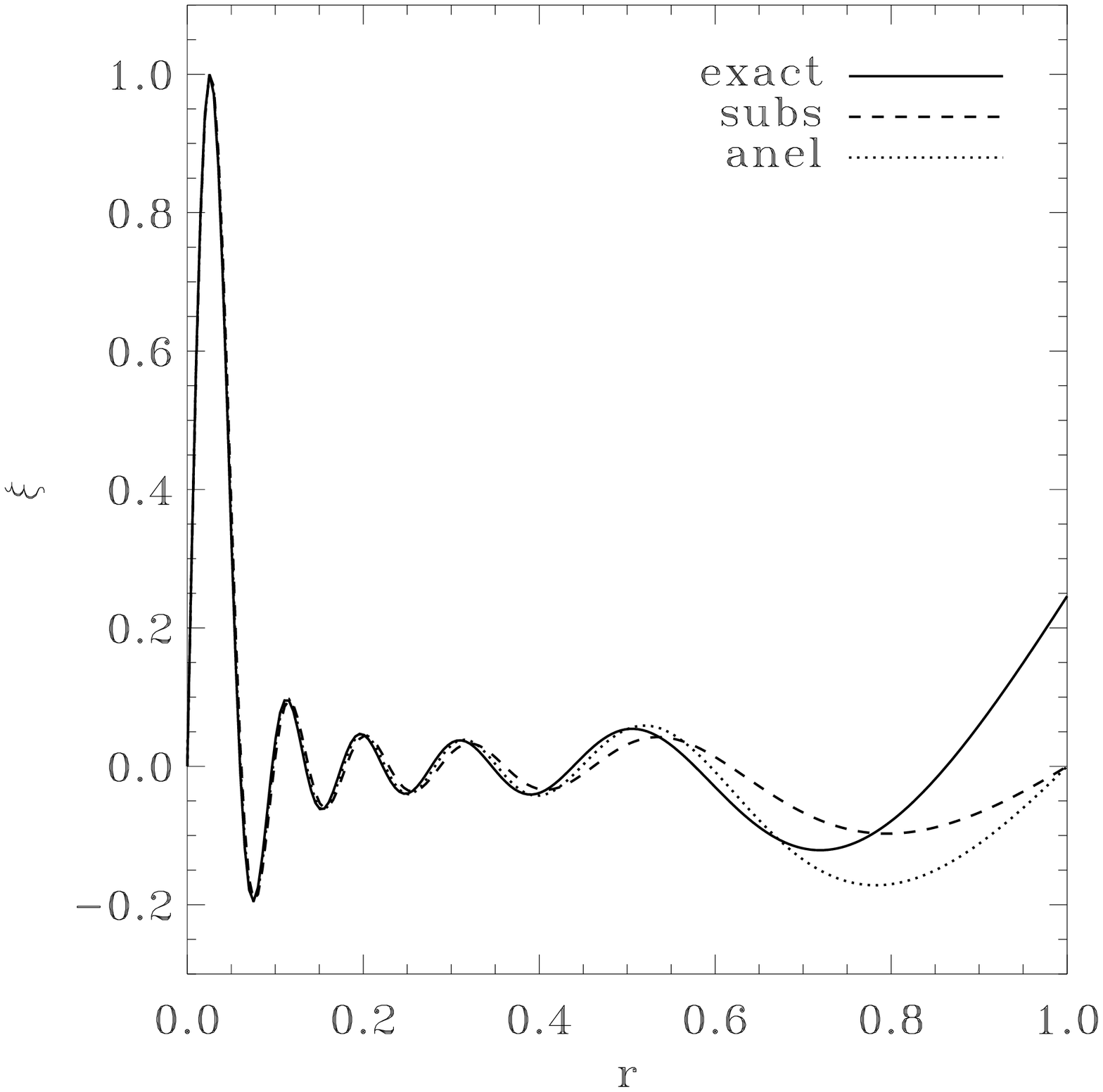}
\includegraphics[width=8cm]{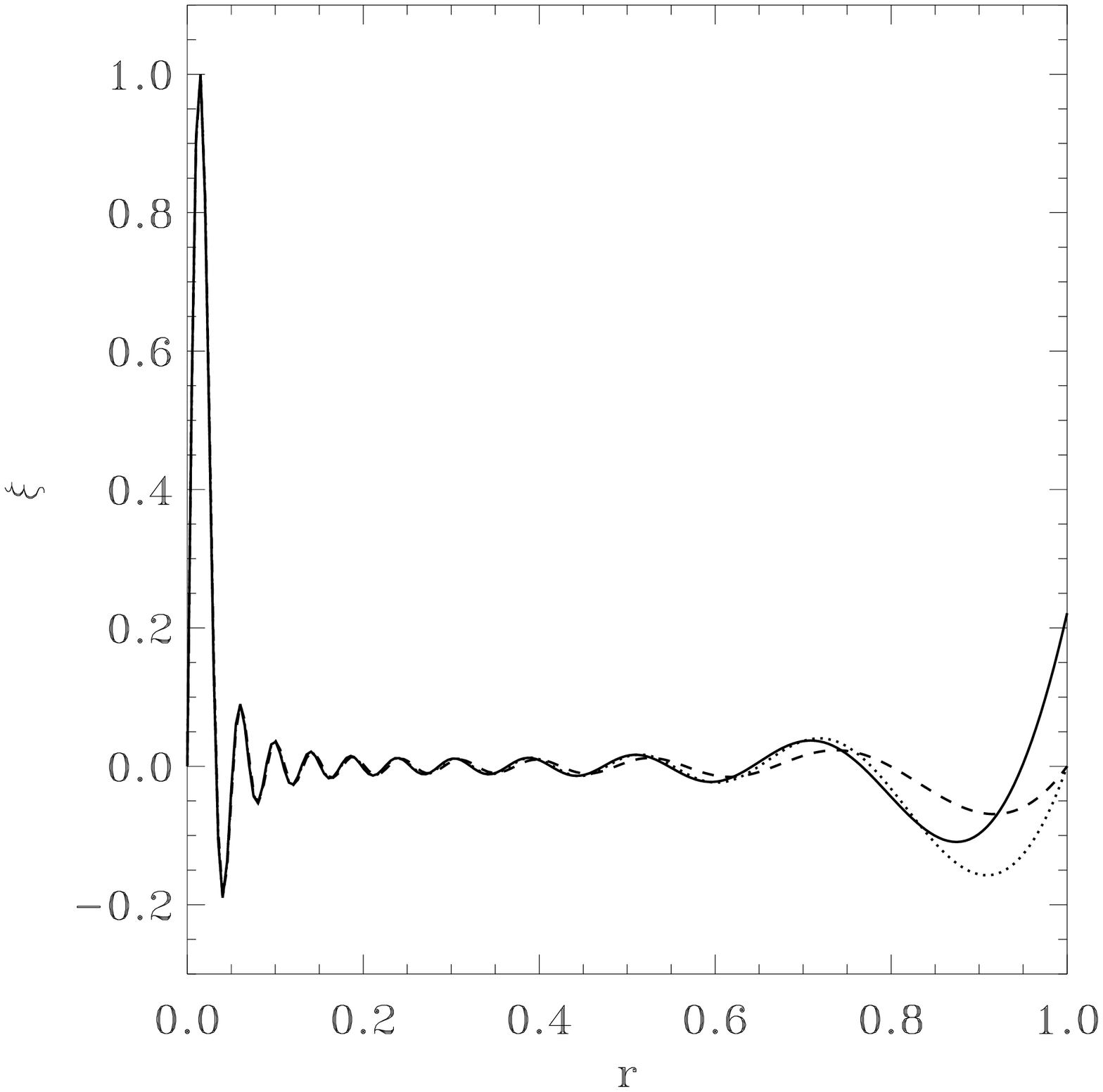}}
\caption[]{Normalized eigenfunctions $\xi$ with $\ell=2$ and $k=10$
(left) and $k=20$ (right) for the polytrope $n=3$. As in Fig.~1, exact
solutions are denoted by solid lines (with $\omega^2_{10} \simeq
1.98\times 10^{-3}$ and $\omega^2_{20} \simeq 5.99\times 10^{-4}$) while
the dashed and dotted lines correspond to their subseismic and anelastic
approximations, respectively (with $\omega^2_{\hbox{\tiny subs}} \simeq
2.08\times 10^{-3}, \ \omega^2_{\hbox{\tiny anel}} \simeq 2.02\times
10^{-3}$ for $k=10$ and $\omega^2_{\hbox{\tiny subs}} \simeq
6.15\times 10^{-4}, \ \omega^2_{\hbox{\tiny anel}} \simeq 6.04\times
10^{-4}$ for $k=20$).} 
\label{poly} 
\end{figure*}

\section{Examples}

As in paper I we consider two polytropes: one of constant density and
one of index $n=3$.

\subsection{The homogeneous star model}

In this case analytic solutions exist either for the exact or the
approximate equations (see paper I).

In the asymptotic case of large wavenumbers ($k\rightarrow\infty$),
one finds that

\[ 
\omega^2 = \Delta-\sqrt{\Delta^2+\ell(\ell+1)} =
-\frac{\ell(\ell+1)}{2\Delta} + \od{\frac{\ell^2(\ell+1)^2}{\Delta^3}},
\]
with $\Delta=\gamma \lc k \lp \ell+k+\frac{5}{2} \rp + \ell +\frac{3}{2}
\rc - 2$. Therefore

\[
\omega^2\simeq-\frac{2\ell(\ell+1)}{\gamma}\frac{1}
{2k(2k+2\ell+5)+4\ell+6-4/\gamma}.
\]

Now using (22) of paper I we find that for the anelastic approximation

\[
\omega^2_{\hbox{\tiny anel}} =
-\frac{2\ell(\ell+1)}{\gamma}\frac{1}{2k(2k+2\ell+5)+4\ell+6},
\]
while the subseismic expression (21) can be rewritten as:

\[
\omega^2_{\hbox{\tiny subs}} =
-\frac{2\ell(\ell+1)}{\gamma}
\frac{1}{2k(2k+3\ell+4+2/\gamma)+6\ell+(2\ell+4)/\gamma}.
\]

It is clear from these three expressions that, for high order modes,
the anelastic approximation
is very close to the exact expression. In fact, one finds that
$(\omega^2-\omega^2_{\hbox{\tiny anel}})/\omega^2 \sim k^{-2}$ while
$(\omega^2-\omega^2_{\hbox{\tiny subs}})/\omega^2 \sim k^{-1}$, that is,
the anelastic expression converges {\it quadratically} while the subseismic
one only {\it linearly}.

This better behaviour of the \AAP\ is confirmed by the shape of the
eigenfunctions as shown in figure~\ref{homo}. There, we clearly see that
the subseismic solution is good only in the central regions ($r< 0.2$)
while the \AAP\ remains close to the exact solution almost to the surface.

\subsection{The polytrope $n=3$}

For a polytrope $n=3$ a similar behaviour exists although the difference
between the two approximations is less pronounced. In I we observed that
the eigenfrequencies converged at different rates, the \AAP\ one
converging faster. Here we plot two eigenmodes of high order ($k=10$ and
$k=20$) computed with the two approximations and with the complete
equations (see figure~\ref{poly}). As expected, while both approximations
describe the central regions very accurately, the anelastic one remains
closer to the `exact' solutions on a larger volume. For the $k=20$-mode,
it departs noticeably from the exact solution close to the surface
($r\sim0.83$).

\section{Conclusion}

\psfrag{0}{0}\psfrag{omegan}{$\omega_N$}\psfrag{omegac}{$\omega_c$}
\psfrag{omega}{$\omega$} \psfrag{eps}{$\eps\omega$}
\psfrag{gravity}{\hspace*{-7.5mm}$\overbrace{\hspace*{2.5cm}}^{\rm
\displaystyle Gravity \; modes}$}
\psfrag{acoustic}{\hspace*{-9.5mm}$\overbrace{\hspace*{2cm}}^{
\parbox{2.5cm}{\rm Acoustic modes\\ rejected to $\infty$}}$}

\begin{figure}
\vspace*{1cm}
\centerline{\includegraphics[width=8cm]{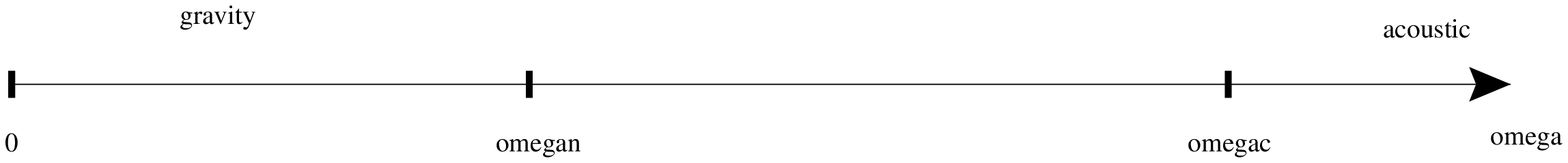}}
\vspace*{1cm}
\psfrag{gravity}{\hspace*{-1.0cm}$\overbrace{\hspace*{1.2cm}}^{
\parbox{2.5cm}{Asymptotic gravity\\ \centerline{modes}}}$}
\psfrag{acoustic}{\hspace*{-1.2cm}$\overbrace{\hspace*{2.5cm}}^{
\parbox{2.5cm}{\rm Acoustic modes}}$}
\centerline{\includegraphics[width=8cm]{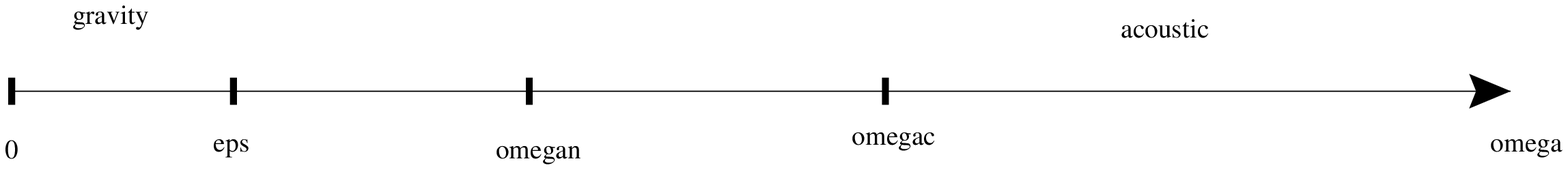}}
\caption[]{A schematic picture of the modes of a star viewed from
the anelastic viewpoint (above) and subseismic viewpoint (below).
$\omega_N$ and $\omega_c$ are respectively the frequency of the lowest
order gravity and acoustic modes.}
\label{schema}
\end{figure}

In this paper we tried to clarify the differences between the
subseismic and anelastic approximations which both aim at describing the
low frequency spectrum. The \SA\ appears when one concentrates on the
low frequency high radial order modes in the central region of a star;
no constraint is imposed to the \BV frequency.

On the other hand the \AAP\ assumes a weak stratification but imposes no
constraint on the degree of the mode.

Hence, while the \AAP\ makes the \BV frequency, and thus the frequency
of all gravity modes, vanishingly small compared
to acoustic frequencies, the \SA\ focuses on gravity modes whose radial
order is very large and hence have small frequencies compared to
acoustic ones.

In other words, the \AAP\ removes the elasticity of the fluid by rejecting
acoustic frequencies to infinity and therefore allows for a description of
the full spectrum of gravity modes while the \SA, keeping $\omega_c$
and $\omega_N$ in a finite ratio, concentrates on one part of the
spectrum, namely that containing high radial order modes which are the
least sensitive to the elasticity of the fluid.  This situation is
summarized in figure~\ref{schema}.

Since in stars the situation is often that $\omega_{N}\ll
\omega_{c}$, the use of the \AAP\ is recommended as it is likely
closer to the solutions of the complete equations; on the other hand,
the \SA\ may be useful when one needs an analytic expression of gravity
modes in the central regions of a star.

Finally, it is worth mentioning the work of \cite{Durr89} who discussed
these two approximations in the context of atmospheric sciences. In this
field, where the \SA\ is called the ``pseudo-incompressible appoximation"
and the anelastic approximation the ``modified anelastic approximation",
the \SA\ appears to be superior to the \AAP\ as it conserves the energy,
a property which is important for nonlinear problems.  This result shows
that the best choice for filtering out acoustic modes is dependent on
the problem at hands.  Therefore, our results which favour the \AAP\
when searching for low-frequency modes of stars, may be specific to
eigenvalue problems.

\section*{acknowledgements}
BD acknowledges support from the European Commission under Marie-Curie
grant no. HPMF-CT-1999-00411.

\label{lastpage}
\end{document}